# Atomic-scale Manipulation of Single-Polaron in a Two-Dimensional Semiconductor

Huiru Liu[1,3]†, Aolei Wang [2]†, Ping Zhang[1,3], Chen Ma[1,3], Caiyun Chen[1,3], Zijia Liu[1,3], Yiqi Zhang[1,3], Baojie Feng[1,3], Peng Cheng[1,3], Jin Zhao [2]*, Lan Chen[1,3,4]*, Kehui Wu[1,3,4]*

[1] *Institute of Physics, Chinese Academy of Sciences, Beijing 100190, China*

[2]*Department of Physics, CAS Key Laboratory of Strongly-Coupled Quantum Matter Physics, and ICQD/Hefei National Laboratory for Physical Sciences at Microscale, University of Science and Technology of China, Hefei, Anhui 230026, China*

[3] *School of Physical Sciences, University of Chinese Academy of Sciences, Beijing, 100190, China*

[4] *Songshan Lake Materials Laboratory, Dongguan, Guangdong, 523808, China*

† These authors contributed equally to this work.

*Emails: khwu@iphy.ac.cn (K.W.); lchen@iphy.ac.cn (L.C.); zhaojin@ustc.edu.cn (J. Z.)

**Abstract**

**Polaron is a composite quasiparticle derived from an excess carrier trapped by local lattice distortion, and it has been studied extensively for decades both theoretically and experimentally. However, atomic-scale creation and manipulation of single-polarons in real space have still not been achieved so far, which precludes the atomistic understanding of the properties of polarons as well as their applications. Herein, using scanning tunneling microscopy, we succeeded to create single polarons in a monolayer**

**two-dimensional (2D) semiconductor, CoCl$_2$. Combined with first-principles calculations, two stable polaron configurations, centered at on-top and hollow sites, respectively, have been revealed. Remarkably, a series of manipulation progresses — from creation, erasure, to transition — can be accurately implemented on individual polarons. Our results pave the way to understand the polaronic physics at atomic level, and the easy control of single polarons in 2D semiconductor may open the door to 2D polaronics including the data storage.**

## Introduction

Polaron is a fundamental physical phenomenon associated with the behavior of charges in insulators or semiconductors. Different from metal where charges can flow as current and easily be compensated, insulator or semiconductor will usually have charges stuck inside. For example, rubbing glass with fur can build up significant charges in both materials. Obviously, understanding the atomistic details of these charge states in insulators or semiconductors is fundamental and necessary. Indeed, the studies on polaron have a long history: In 1932, Landau discussed the idea that free electrons in ionic lattice would be trapped at a substantially distorted region [1]. The concept of polaron was formally proposed by Solomon Pekar in 1946 [2]. It describes that an electron introduced into the conduction band of an ionic crystal may cause a local deformation in the lattice due to the long-range Coulomb interaction, which results in the formation of local potential that blocks the escape of the electron. Since then, tremendous experimental observations and theoretical models have been established. Theoretical models of polarons were mainly based on the seminal works of Herbert Frohlich [3,4] and Theodore Holstein [5,6] in 1950s, which provided a self-consistent picture of polaron and had been followed ever since. Recently, polarons have been identified in a large varieties of materials, such as alkali chloride KCl [7], transition metal oxides $TiO_2$ [8-11], metal halide perovskites [12,13], organic semiconductors [14], transition metal dichalcogenides (TMD) [15] and their heterostructures [16,17].

Polaron has significant influences on the properties of semiconducting materials, such as lattice reconstructions [18], carrier mobility [19,20], surface adsorption and catalysis [8,21-23], ferromagnetic transition [9,24,25], superconductivity [26-29] and other many-body correlation states [30-32]. The investigations and applications of polaron have has been recently developed as an emerging field in nanotechnology, named as 'Polaronics'. In previous investigations, polarons have been experimentally accessed by various techniques such as Raman [33-35], THz spectroscopy [36], angle-resolved photoelectron emission spectroscopy [37], electron paramagnetic resonance [38], and scanning tunneling microscopy (STM) [25,39,40]. However, the capability to view, create and manipulate single polarons in real space still remains elusive, which limits the understanding and application of polaronics at atomic scale. Although the possibility of data storage by writing and erasing single polaron in materials have been theoretically proposed by Bondarenko *et al.* based on kinetic Monte Carlo simulations [41], relevant experiments have not been realized so far.

STM is an ideal tool for probing and manipulating local objectives with atomic precision [42]. In this work, we studied single polarons in transition metal dichloride $CoCl_2$, an intrinsic magnetic 2D semiconductor, by STM. Through applying a voltage pulse, we can inject an electron into the monolayer $CoCl_2$ and created a single electron polaron. It was found that the polarons have two stable forms, labeled as type-I and type-II polarons. First principles calculations indicate that they can be described as spin-polarized electrons bonded to lattice distortions centered at either on-top or hollow sites, respectively. Moreover, the creation and erasure of single polarons, as well as transition between type-I and type-II polarons, can be realized by precisely regulating the voltage pulse. Both types of polarons falls into the small polaron category [43], and are single spin information carriers. Our work paves a way to polaronics based on 2D materials.

## Results & discussion

**Growth and characterization of CoCl$_2$ monolayer.** CoCl$_2$ is a layered material stabilized by weak interlayer van der Waals (vdW) interaction [44], and the atomic structure of CoCl$_2$ monolayer is shown in Fig.1d. Each CoCl$_2$ monolayer consists of a sub-layer of Co atoms sandwiched between two Cl sub-layers. In our experiment, the CoCl$_2$ monolayer was grown on highly oriented pyrolytic graphite (HOPG) substrate. The STM image in Fig. 1a reveals that the monolayer CoCl$_2$ grows in a dendritic structure starting from step edges of HOPG. The height of monolayer CoCl$_2$ is 480±5 pm, clearly larger than that of HOPG (340±5 pm), as illustrated in Fig. 1a. Such dendritic growth mode can be described in the diffusion limited aggregation framework [45]. The atomic-resolution STM images of the monolayer CoCl$_2$ surface shown in Fig. 1b and 1c reveal a triangular lattice with periodicity of 354±2 pm, consistent with the lattice parameters (354.5~355.3 pm) for CoCl$_2$ single crystal [46,47] and powder [48]. Moreover, depending on the crystallographic orientation of the CoCl$_2$ monolayer with respect to the HOPG substrate, different moiré patterns can be found on the surface of CoCl$_2$ monolayer. For examples, Figure 1c exhibits a moiré periodicity of about 1.18 nm, whereas there is no moiré modulation in Fig. 1b.

The low temperature (4K) scanning tunneling spectroscopy (STS) measurements taken on the monolayer CoCl$_2$ reveal typical semiconducting band feature with bandgap ~1.7 eV, as shown in Fig. 1e. The Fermi level is closer to the edge of conduction band, suggesting the monolayer CoCl$_2$ is electron-dopped. Within the band gap, the d$I$/d$V$ curve exhibits V shape (inset of Fig. 1e), corresponding to the electronic structure of underlying HOPG. For comparison, we calculated the band structure of free-standing monolayer CoCl$_2$ by DFT with HSE functional. To reveal the intrinsic band structure of monolayer CoCl$_2$, the HOPG substrate is not included in the calculation. As seen in Fig. 1f, the calculated band structure has a large bandgap (4.1 eV), which is however much larger than the experimental value and suggests a significant influence of the HOPG substrate. Indeed, the calculated band structure of monolayer CoCl$_2$ including the HOPG substrate exhibits emerging states in the bandgap, which is consistent with experiment, as will be discussed later.

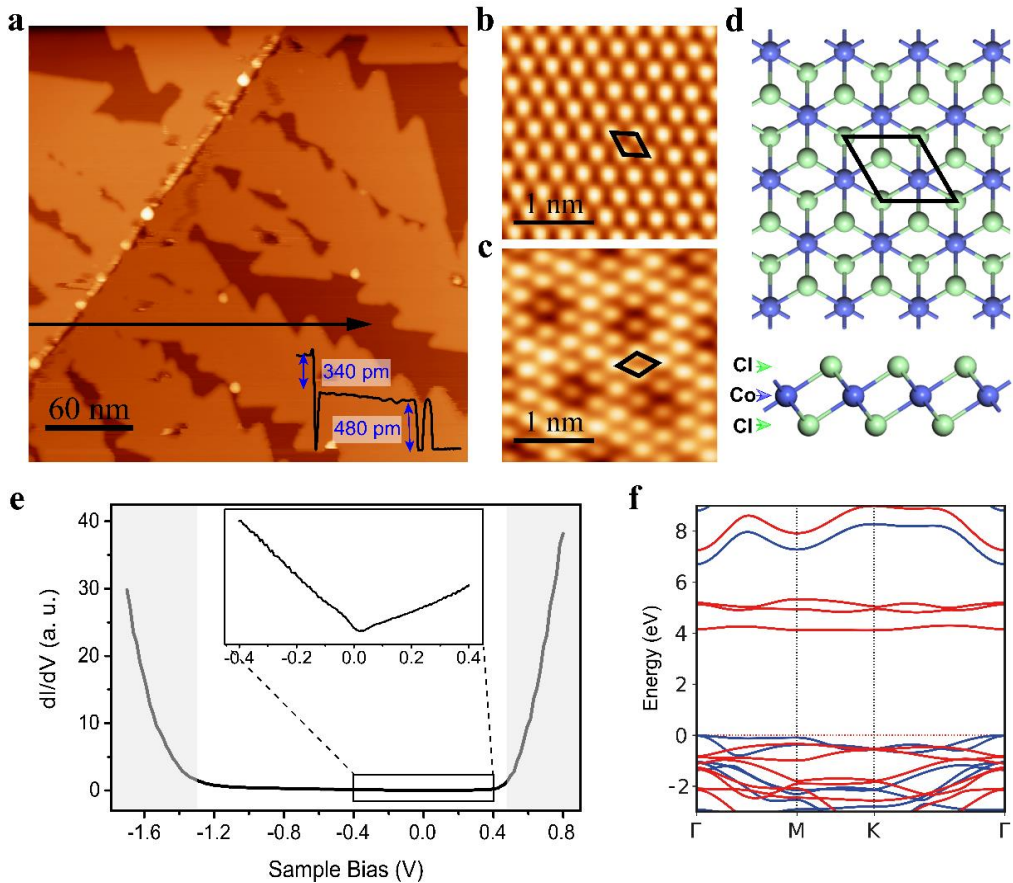

**Fig.1 The STM/S characterizations of CoCl$_2$ monolayer on HOPG substrate.** **(a)** Large-area STM image ($V_s = -1$ V, $I = 15$ pA) of CoCl$_2$ with sub-monolayer coverage. Inset: line profile along the black arrow across steps of HOPG and CoCl$_2$. **(b, c)** High resolution STM images of CoCl$_2$ surface. The protrusions correspond to Cl atoms, **b** exhibits no moiré pattern ($V_s = 750$ mV, $I = 20$ pA) while **c** exhibits moiré periodicity of 1.18 nm ($V_s = -2$ V, $I = 40$ pA) due to different orientations of the CoCl$_2$ monolayer with respect to the HOPG substrate. **(d)** Top and side views of the atomic structural model of CoCl$_2$ monolayer. Purple and green balls represent Co and Cl atoms, respectively. The 1×1 unit cell is labeled by black rhombus in **b-d**. **(e)** d$I$/d$V$ curve taken on CoCl$_2$ surface revealing the typical semiconducting nature. The inset shows the d$I$/d$V$ curve within small bias range around the Fermi level [-0.4 eV, +0.4 eV]. **(f)** Calculated band structure of the CoCl$_2$ by DFT with HSE functional, exhibiting a band gap of 4.1 eV. Blue and red lines represent spin-up and spin-down electron bands, respectively. The reference energy refers to valence band maximum energy.

**Manipulation of single polarons.** Interestingly, we found a highly reproducible hysteresis phenomenon when performing I-V measurements on the CoCl$_2$ monolayer. Sweeping the sample bias between -1.2 V and +1.1 V, very often we encounter sudden current jumps in either side around ±1.0 V, where the smoothly increasing current suddenly become smaller (Fig. 2a). We first consider the jump at positive bias side. As positive sample bias corresponds to electron injection from tip to sample, we infer that the injecting electrons may stimulate a critical process when the bias voltage is above a threshold. To show the effect more clearly, we applied a positive bias of 0.8 V with the feedback loop off, and monitored the tunneling current within a time window. Very often a current jump from high current state to a low current state has been recorded, as indicated in Fig. 2b. Consequently, a dramatic ring-like feature will appear in the d$I$/d$V$ map taken afterwards at the same position.

We then consider the current jump in the negative bias side, and we found that it corresponds to the annihilation of the ring feature. As shown in Fig. 2c, when applying a negative bias of -1.0 eV with holding the STM tip on an existing ring feature, within a time window very often a current jumping from high current state to a low current state will be recorded. Consequently, we find that the ring-like feature has been annihilated from the surface (panel 3 in Fig. 2c).

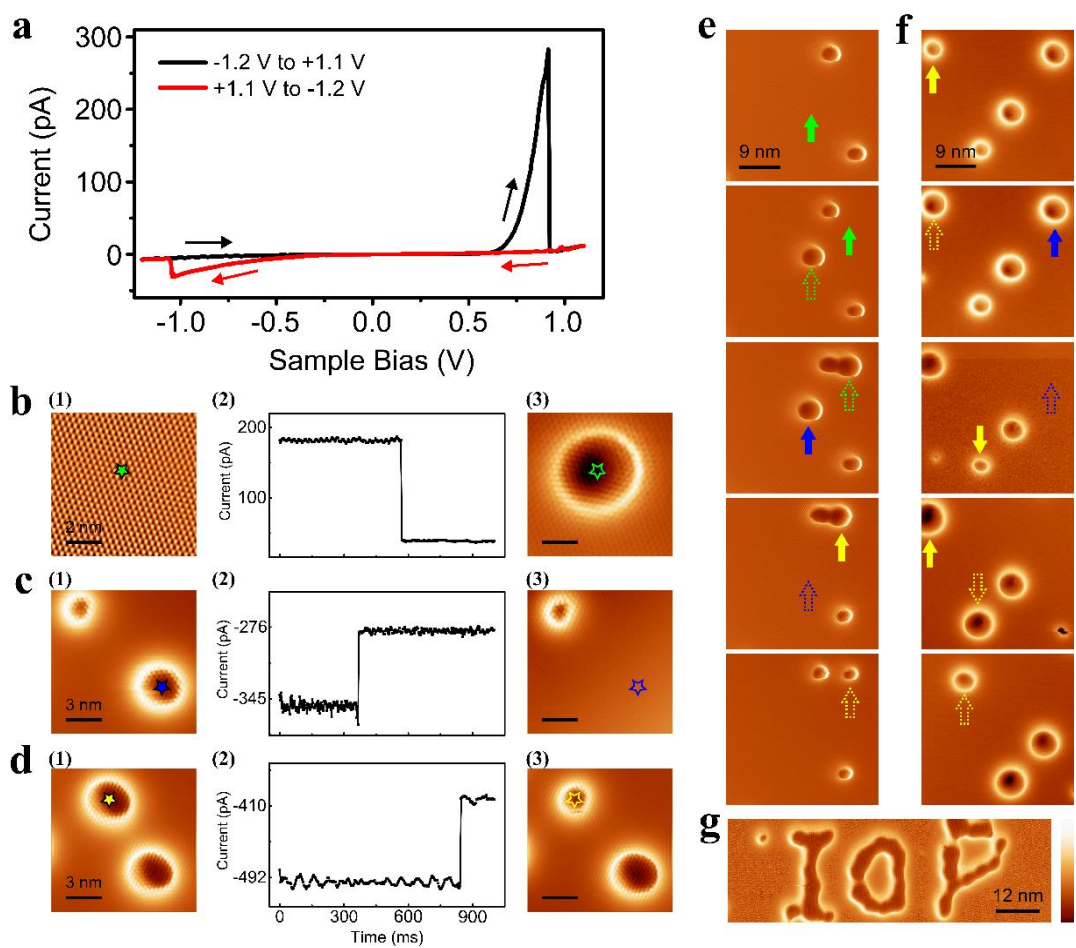

**Fig. 2 Manipulation process of single polarons.** (**a**) Typical I-V curve sweeping between -1.2 V to +1.1V measured on defect-free region of monolayer $CoCl_2$ surface, showing two current jumps. (**b-d**) The process of a single operation event by applying bias pulses. (**b**) a voltage pulse of 800 mV applying at a clean area. (**c**) a voltage pulse of -1.0 V applying at center of one ring. (**d**) a voltage pulse of -1.2 V applying at center of the bigger ring. The left panel (**1**) are the d$I$/d$V$ maps of an area before event, and the right panel (**3**) are the d$I$/d$V$ maps of same area after event. The middle panel (**2**) are the recorded current change during voltage pulsing. (**e-f**) Two series of d$I$/d$V$ maps, from top panel to bottom panel, showing successive manipulation events. The green, yellow and blue arrows represent writing, transition and erasure events, respectively. (**g**) A d$I$/d$V$ map showing an 'IOP' pattern which has been created by multiple writing processes. The scanning parameters of all d$I$/d$V$ maps are: $V_s$ = 750 mV, $I$ = 5 pA, 90 nm × 23 nm, and with a same tip.

Carefully inspecting the ring-like features, we find there are two types of rings, smaller one and bigger one, as shown in Fig. 2c and 2d. Both of them can be created or erased with bias pulse as discussed above. Moreover, the transition between big and small rings can also be realized by precisely regulating the voltage pulse (yellow arrows in Fig. 2e and 2f). We can also implement all the three kinds of operations sequentially on a single ring. Two examples are shown in Figs. 2e and 2f, where the green, blue, and yellow arrows represent writing, erasure, and transition operations, respectively. As a demonstration of our capability to manipulate individual ring features, an 'IOP' pattern consisting of 40 rings has been created by successive writing individual rings (Fig. 2g), which means that one can design and implement any artificial patterns with proper manipulation and also suggest the possible application in data storage.

**Evidences of electron polarons.** In order to understand the nature of the ring-like features in monolayer CoCl$_2$, high-resolution STM measurements with various bias have been performed on a bigger ring we created. As shown in Fig. 3a, the STM image and d$I$/d$V$ map taken at positive bias of 740 mV both show a depression at the center of ring with a radius of about 3.16 nm, as deduced from the red dotted curve in Fig. 3b. On the other hand, STM image of same area taken at $V_b$ = −500 mV (down panels of Fig. 3a) reveals only a slightly brighter region, while the clear atomic resolution in this image indicates there is no defect around the ring site. In addition to the dramatic difference of STM images with bias polarity, the bias voltages dependence with the same polarity is also significant. As shown in Fig. 3c, the radii of the features in both the STM images (up panel) and d*I*/d*V* maps (down panel) gradually shrink as the bias increases. Such shrinking features with bias in STM/STS images have been typically observed on charged point defects, single dopants, and adatoms on semiconductor surfaces, like GaAs [49], InP [50], $MoS_2$ [51] and black phosphorus [15]. Therefore, the feature we observed here is very likely caused by local charge due to the screening effect. To further confirm this point, STS were measured along a line across the feature. As shown in Fig. 3d, the conduction band is bended upward around the feature, suggesting the depression and ring shrinking is the result of conduction band bending when approaching the charge center. All the above observations point to the existence of electron polaron – a defect-less lattice with locally trapped electron.

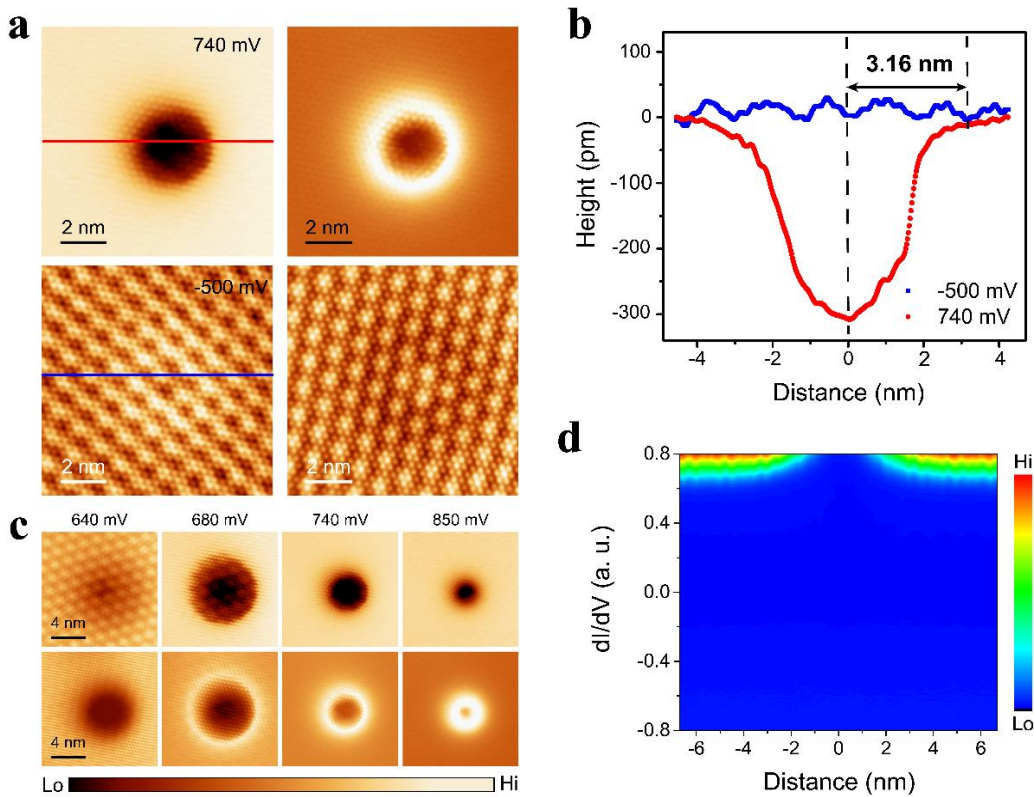

**Fig. 3 The bias-dependent feature of polaron in STM images. (a)** Atomic resolution STM images (left panels) and d*I*/d*V* maps (right panels) of the same individual polaron taken at negative bias -500 mV (up panel) and positive bias 740 mV (down panel). Scale bar: 2 nm. **(b)** Height profiles along the blue and red

lines across the feature in (a), respectively. **(c)** STM images (upper panels) and corresponding d*I*/d*V* maps (lower panels) taken at positive bias range from 640 mV to 850 mV with the same tunneling current $I = 25$ pA. **(d)** The color map consisting of spatially dependent d*I*/d*V* spectroscopy along a line across one ring feature. A significant upward bending of conduction band at the position closer to the ring center was observed.

**Two types of electron polarons.** As we mentioned above, the polarons we created in $CoCl_2$ monolayer exhibit two types with different radius in STM images, which we named as type-I and type-II polarons below. Figure 4a displays the two types of polarons in the same d*I*/d*V* map at 750 meV, clearly showing the ring feature with different radius. The typical radius derived from the line profiles across the polarons, as shown in Fig. 4b, are 3.83 nm (type-I) and 2.97 nm (type-II) respectively. To precisely determine the position of the polarons with respect to the $CoCl_2$ lattice, the atomic model of $CoCl_2$ is superimposed on the d*I*/d*V* map. We found that the centers of two types of polarons are located at different lattice sites: Type-I polaron is centered on chlorine atom site (top site), whereas that of type-II polaron is located at the cobalt atom site (hollow site), as indicated by the blue points. Additionally, in our experiments, type-I polarons are more common than type-II polarons, suggesting that type-I polarons should be slightly more stable than type-II in energy.

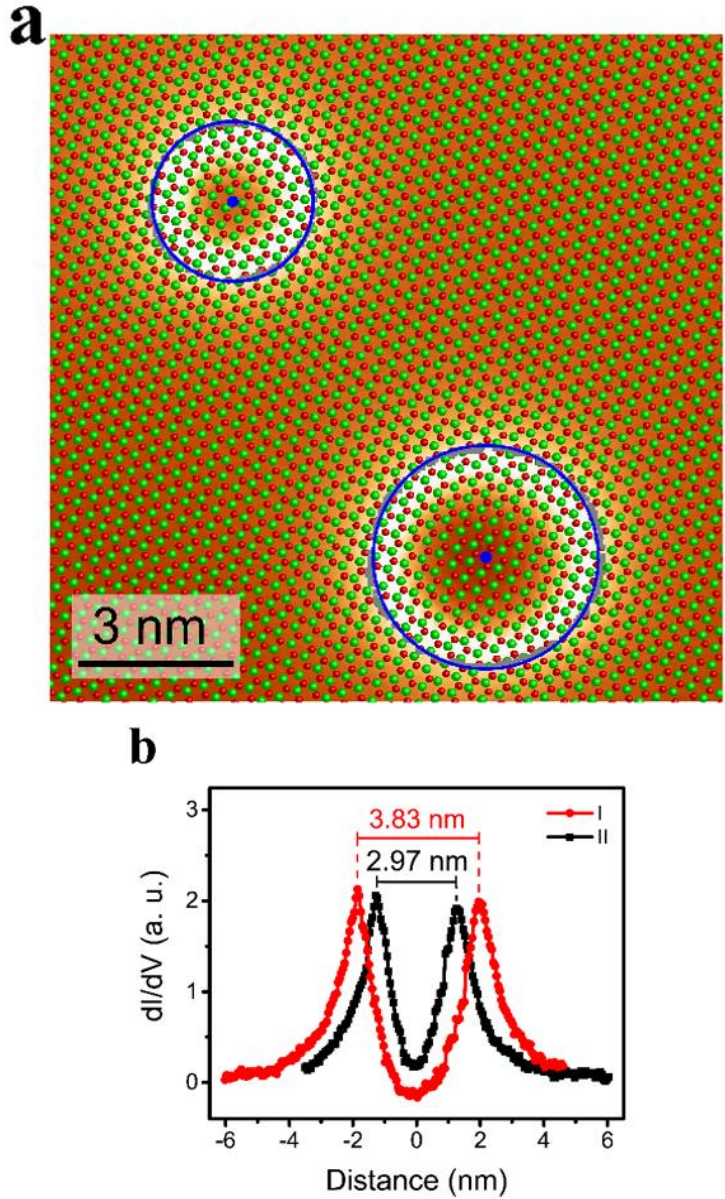

**Fig. 4 Two types of polarons in CoCl$_2$.** **(a)** d$I$/d$V$ maps of monolayer CoCl$_2$ with coexisting type-I (bigger) and type-II (smaller) polarons. Scanning parameter: $V_s = 750$ mV, I = 10 pA. The atomic lattice of monolayer CoCl$_2$ is superimposed on the image, where cobalt and chloride atoms are red and green balls, respectively. The gray points are raw data extracted from isosurface dI/dV = 1.0 ps, while the blue solid lines are corresponding circle curves obtained by fitting. The blue points mark the locations of the respective geometric center of the rings. **(b)** line profile along the lines across the two polarons in (a), showing different radius of the two polarons.

**Theoretical modelling of polarons.** In order to understand of the formation and manipulation mechanism of polarons in CoCl$_2$ monolayer, we performed first-principles calculations, and the results are shown in Fig. 5. By adding an extra electron into the conduction band of monolayer CoCl$_2$, two stable configurations of electron- polaron centered at Cl atom and Co atom sites respectively, were obtained after structure relaxation, nicely corresponding to the type- I and type-II polarons observed in experiments. Figure 5a and 5b give the lattice distortion in CoCl$_2$ for the two types of polarons, and we found the formation of both polarons still preserves the triplet symmetry of system and the magnitude of lattice distortion is in the sub-angstrom range, which matches to that of typical polaron-induced lattice distortions [52]. Note that such small lattice distortion is beyond the spatial resolution limit of STM. Importantly, the corresponding polaron charge density distributions were calculated and displayed in Fig. 5c and 5d. One of them has the charge density evenly distributed among three cobalt atoms around the central Cl atom, whereas the other one has the charge concentrated on a single cobalt atom. These are in good agreement with the experimental observation of two types of polarons located at top site and hollow site of Cl atoms on surface. The binding energies ($E_p$) for two types of polarons are 0.31 eV and 0.22 eV, respectively, indicating the divergent type-I polarons are more stable in energy than the concentrated type-II, which is in line with the experimental statistics.

Moreover, Figure 5e and 5f show that both type-I and type-II polarons host spin-polarized density of states (DOS). For type-I polaron, a localized polaron state is formed at 0.49 eV below the conduction band minimum (CBM). For type-II polaron, the charge trapping by lattice distortion affects the electronic structure more distinctly. Two polaron-induced peaks are formed from 0 to 0.38 eV below the CBM, in addition, the valence band maximum (VBM) is lifted up by around 1 eV. For both types, each polaron carries a spin-polarized electron, suggesting it behaves as a spin information carrier. Based on the calculated charge density, structure distortion, and binding energy, both types of polarons observed in this work can be described by the concept of small polaron in ionic crystal.

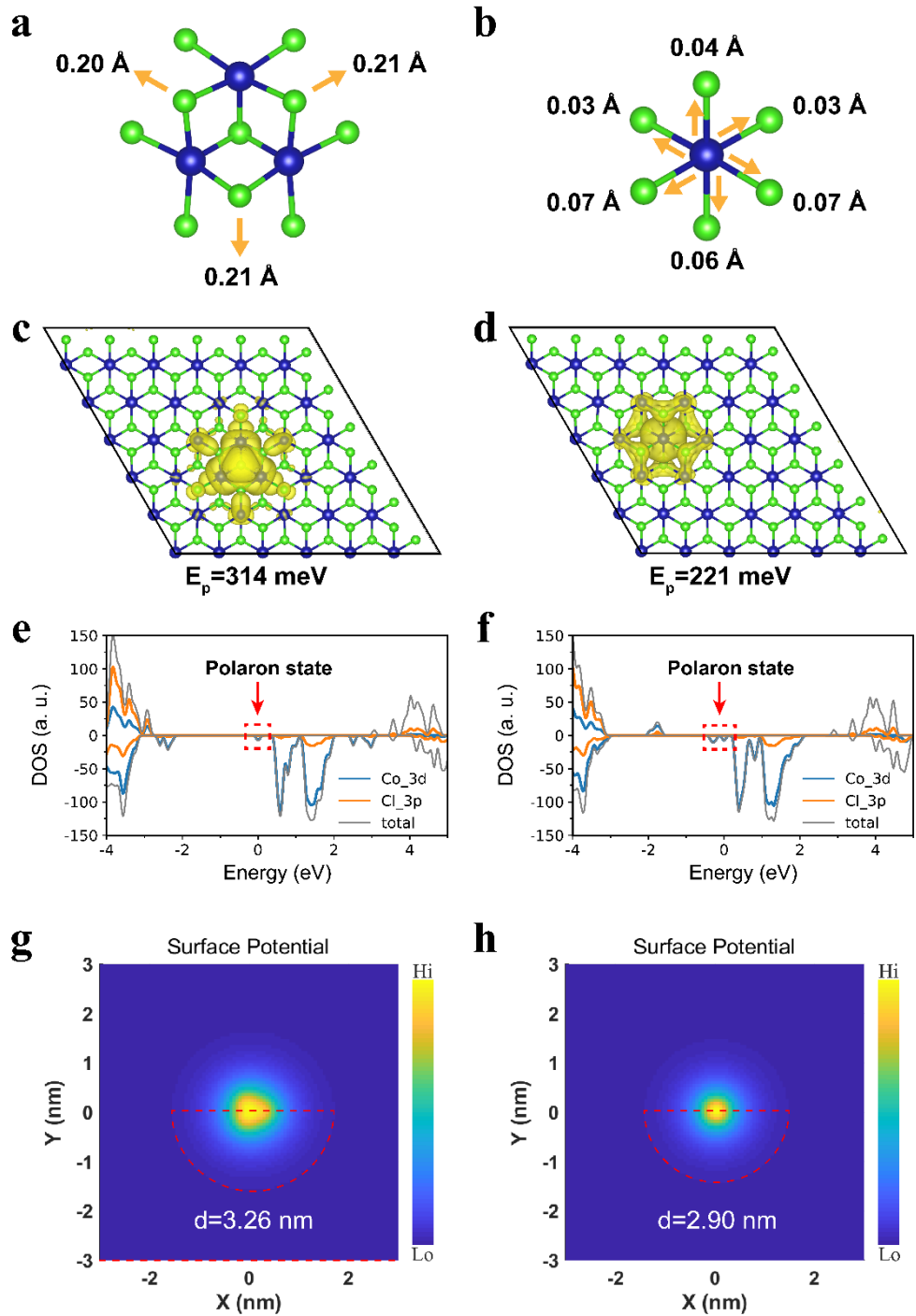

**Fig. 5 Modeling two types of polarons in CoCl$_2$ based on theoretical simulations.** The left (a, c, e, g) and right panels (b, d, f, h) correspond to calculations of type-I and type-II polarons, respectively. **(a, b)** Structure distortion diagrams of monolayer CoCl$_2$ induced by two types of polarons. The arrows represent the directions of atom distortion. Purple and green balls represent cobalt and chlorine atoms, respectively. **(c, d)** The charge density distribution of two types of polarons. The electron in type-I polaron is evenly distributed among three cobalt atoms in the nearest vicinity, whereas electron in type-II polaron is concentrated on a single cobalt atom. The formation energy of type-I polaron (0.314 eV) is higher than type-II polaron (0.221 eV). **(e, f)** Density of states (DOS) of monolayer CoCl$_2$ with single polarons. The reference energy is Fermi energy. The red dotted rectangle marks the in-gap state of polaron. **(g, h)** Simulated surface electrostatic potential diagrams of the charge distribution around type-I and type-II polarons. The solid red fan frames mark 1/2 circle, and the corresponding diameters are labelled.

Considering the depression and ring-like feature of polaron in STM images and d$I$/d$V$ maps corresponding to the conduction band bending due to charge screening effect, the different radius of polarons should be related to the different charge distribution in them. To prove this, we calculated the distribution of electrostatic potential around the two types of polarons (the method is shown in Part I of Supplemental Materials), and the results are shown in Fig. 5g and 5h. The simulated feature diameters are 3.26 nm and 2.90 nm for two types of polarons respectively, which are qualitatively in good agreement with experiments. The smaller values compared with the experiments may be due to overestimation of the screening effect of HOPG substrate in simulations.

**Mechanism of the polarons manipulation.** To find out the mechanism of polaron manipulation, we statistically analyzed the writing probability as a function of the pulse voltage (in our experiments, the writing probability is defined as ratio of the number of writing events to the number of applied pulses (> 100 times) with same energy, current and duration time), as shown in Fig. 6a. When the energy of tunneling electrons is low enough, the writing probability remains almost zero. With the increase of pulse voltage above 800 mV, the ratio of writing events rapidly increases to one, which means nearly 100% possibility of writing for every bias pulsing when electron energy exceeds the threshold value. The Boltzmann fitting can give a threshold voltage about 858 meV at tunneling junction height ($V_s = 600$ mV, $I = 25$ pA).

The writing process is directly related to the injection of electron to the conduction band of $CoCl_2$ monolayer on HOPG. We calculated the DOS of $CoCl_2$ monolayer involving the HOPG substrate. As shown in Fig. 6b, the $CoCl_2$ monolayer keeps its semiconducting property, while the DOS inside the band gap is mostly from the contribution of HOPG. The conductive band edge of $CoCl_2$, which is derived from the Co $3d$ orbitals, is about 600 meV above the Fermi level. The calculated energy position of conductive band edge of $CoCl_2$ (600 meV) is smaller than the experimental writing threshold energy (858 meV). The larger threshold voltage in experiment may be due to the formation of the double barrier in the tunneling junction: one is between tip and the surface of $CoCl_2$ monolayer, and the other between the upper and lower surface of $CoCl_2$ monolayer. The actual bias voltage dropped between tip and $CoCl_2$ monolayer is thus smaller than the bias voltage applied to the tunneling junction.

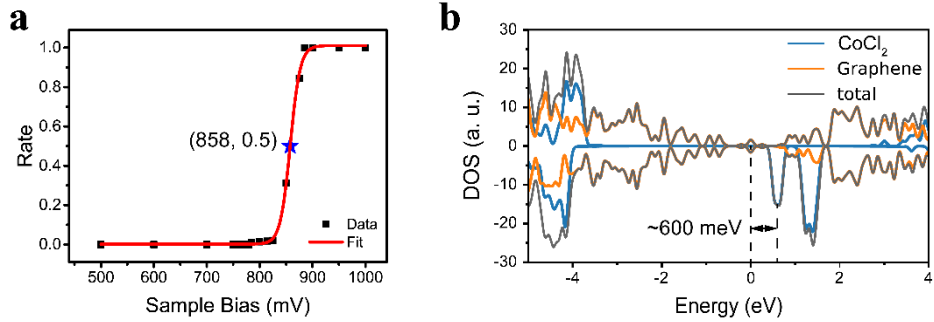

**Fig. 6 The manipulation mechanism of polaron.** (**a**) The statistics of writing rate of polaron with sample bias. The rate is defined as the ratio of the number of successful writes to the total number of operations with the same condition. The solid red curve is the result of Boltzmann fitting, which gives the threshold bias of 858 meV with pulse width of 1 s at tunneling parameter $V_s = 600$ mV, $I = 25$ pA. (**b**) The DOS of $CoCl_2$ monolayer on four layers of graphene. The edge of $CoCl_2$ conduction band is about 600 meV above Fermi level, which is close to the experimental threshold bias of writing.

## Conclusion & Perspective

We discovered electron polarons on a two-dimensional semiconductor, $CoCl_2$ monolayer on HOPG substrate. Systematic investigations were performed to characterize the polarons, including the location of the lattice distortion, the charge screening, and band bending. Remarkably, for the first time, single polaron can be created and manipulated at atomic level with the STM tip. Note that controllable charge states has only been realized in surface defect or adatoms where the binding energies are much larger. Each polaron is a single spin carrier, and the controllable polarons in $CoCl_2$ monolayer provide us an ideal system for both the study of polaron physics in atomic scale, and the applications in data storage, spintronics and quantum computing. This work paves a new pathway to the realization of polaronic devices.

**Methods**

**Sample fabrication and STM/S characterization**

Our experiments were carried out in a home-built low-temperature STM/MBE system with a base pressure better than $5 \times 10^{-11}$ Torr. A clean HOPG substrate was prepared by mechanical exfoliation in air to obtain a fresh surface and then quickly introduced into UHV chamber, and annealing at 900 K. The $CoCl_2$ monolayer were prepared by directly evaporating anhydrous $CoCl_2$ beads (Sigma, 99.999%) on HOPG substrate kept at room temperature. After growth, the sample was transferred to the STM chamber and measured with a tungsten tip under liquid-helium temperature (4K). The STS were measured by a lock-in technique, in which an ac voltage of 20 mV and 659 Hz was superimposed on the given sample bias. The STM images and d$I$/d$V$ maps were obtained at constant-current mode.

**Theoretical calculation based on DFT**

The geometry optimization and electronic structure of polarons were performed with the quickstep module of the CP2K program package [53,54] within the Gaussian and Plane Waves (GPW) framework. HSE06 [55,56] exchange-correlation functional together with Goedecker-Teter-Hutter (GTH) pseudopoetentials [57] was applied. The cutoff and relative cutoff energies of the auxiliary plane wave basis sets were converged to energy differences smaller than $10^{-6}$ hartree/atom. Triple-zeta "MOLOPT" basis sets [58] were used for Co and Cl. The $CoCl_2$ monolayer was modeled using a hexagonal $6 \times 6$ supercell with 108 atoms sampled at the Γ point. A vacuum space larger than 25 Å was adopted to avoid any interaction between two adjacent slabs.

The band structure of $CoCl_2$ monolayer and DOS of $CoCl_2$/HOPG were preformed using the Fritz-Haber-Institute ab initio molecular simulations (FHI-aims) package [59-61]. A scalar relativistic treatment with the atomic ZORA approximation [59] was included in calculations. We used the "light" setting for numerical atom-centered orbital basis sets in FHI-aims. In our calculations, the $CoCl_2$/HOPG system was built by depositing a 2×2 supercell of monolayer $CoCl_2$ on a 3 × 3 supercell of 4 layers' graphene. The lattice mismatch between the $CoCl_2$ and graphene substrate in the constructed heterostructure was only about 4%. The vdw interaction functional using the method of Tkatchenko-Scheffler method with iterative Hirshfeld partitioning [62] was employed in the vdW heterostructure calculations. The structures were relaxed using a Broyden-Fletcher-Goldfarb-Shanno (BFGS) optimization algorithm until the maximum force on each atom was less than 0.01 eV/Å. The convergence criteria of $10^{-6}$ eV for the total energy of the systems were used.

**Acknowledgements**

This work was financially supported by the National Key R&D Program of China (2021YFA1400500, 2018YFE0202700), National Natural Science Foundation of China (11825405, 12134019, 11974322, 12125408), Beijing Natural Science Foundation (Z180007), the Strategic Priority Research Program of the Chinese Academy of Sciences (XDB30000000).

**Author Contributions:**

KW and LC designed supervised the project. HL performed experiments and data analysis. AW perform first-principles calculations under the supervision of JZ. HL and KW prepared the manuscript with contributions from LC, AW and JZ. All other authors contributed to the experimental setup and discussion during the research in this project.